\documentclass[graybox]{svmult}

\usepackage{mathptmx}       
\usepackage{helvet}         
\usepackage{courier}        
\usepackage{type1cm}        
\usepackage{makeidx}         
\usepackage{graphicx}        
\usepackage{multicol}        
\usepackage[bottom]{footmisc}

\usepackage{bm}
\usepackage{latexsym}
\usepackage{dcolumn}
\usepackage{amsmath,amsfonts,amssymb}
\usepackage{graphicx,epsfig}
\usepackage{subfig}
\usepackage{ragged2e}
\usepackage{caption}
\usepackage{color}

\def\beq{\begin{equation}}
\def\eeq{\end{equation}}
\def\bea{\begin{eqnarray}}
\def\eea{\end{eqnarray}}
\def\benu{\begin{enumerate}}
\def\eenu{\end{enumerate}}
\def\nn{\nonumber}

\def\l{\left}
\def\r{\right}

\newcommand{\chf}[1]{\texttt{\footnotesize{#1}}}

\reversemarginpar

\def\DM{\mathrm{d}}
\def \mysymb {\Delta}

\newcommand\blfootnote[1]{%
  \begingroup
  \renewcommand\thefootnote{}\footnote{#1}%
  \addtocounter{footnote}{-1}%
  \endgroup
}

\newenvironment{rcases}
  {\left.\begin{aligned}}
  {\end{aligned}\right\rbrace}

\makeindex             

\begin{document}

\title*{Accelerated Observers, Thermal Entropy, and Spacetime Curvature}
\author{Dawood Kothawala}
\institute{
Dawood Kothawala \at Indian Institute of Technology Madras, Chennai, \email{dawood.ak@gmail.com}
}
%
%
\maketitle

\abstract*{Assuming that an accelerated observer with four-velocity ${\bm u}_{\rm R}$ in a curved spacetime attributes the standard Bekenstein-Hawking entropy and Unruh temperature to his ``local Rindler horizon", I show that the {\it change} in horizon area under parametric displacements of the horizon has a very specific thermodynamic structure. 
Specifically, it entails information about the time-time component of the Einstein tensor: ${\bm G({\bm u}_{\rm R}, {\bm u}_{\rm R})}$. Demanding that the result holds for all accelerated observers, this actually becomes a statement about the full Einstein tensor, ${\bm G}$. 
\newline\indent
I also present some perspectives on the free fall with four-velocity ${\bm u}_{\rm ff}$ across the horizon that leads to such a loss of entropy for an accelerated observer. Motivated by results for some simple quantum systems at finite temperature $T$, we conjecture that at high temperatures, there exists a universal, system-independent curvature correction to partition function and thermal entropy of {\it any} freely falling system, characterised by the dimensionless quantity $\Delta = {\bm R({\bm u}_{\rm ff}, {\bm u}_{\rm ff})} \l(\hbar c/kT\r)^2$.
}

\abstract{Assuming that an accelerated observer with four-velocity ${\bm u}_{\rm R}$ in a curved spacetime attributes the standard Bekenstein-Hawking entropy and Unruh temperature to his ``local Rindler horizon", I show that the {\it change} in horizon area under parametric displacements of the horizon has a very specific thermodynamic structure. 
Specifically, it entails information about the time-time component of the Einstein tensor: ${\bm G({\bm u}_{\rm R}, {\bm u}_{\rm R})}$. Demanding that the result holds for all accelerated observers, this actually becomes a statement about the full Einstein tensor, ${\bm G}$. 
\vspace{.25cm}
\\
I also present some perspectives on the free fall with four-velocity ${\bm u}_{\rm ff}$ across the horizon that leads to such a loss of entropy for an accelerated observer. Motivated by results for some simple quantum systems at finite temperature $T$, we conjecture that at high temperatures, there exists a universal, system-independent curvature correction to partition function and thermal entropy of {\it any} freely falling system, characterised by the dimensionless quantity $\Delta = {\bm R({\bm u}_{\rm ff}, {\bm u}_{\rm ff})} \l(\hbar c/kT\r)^2$.
}

\section{Gravity and Thermodynamics}
\label{sec:1}
It has been well known for a long time that statistical mechanics in presence of gravitational interactions exhibits several peculiar features \cite{lynden-bell-paddy}, deriving mostly from the fact that gravity couples to {\it everything}, and operates {\it unshielded} with an {\it infinite range}. Many of these peculiarities, such as {negative} specific heat, however, attracted attention {only after} they were encountered in the context of {black holes}. Indeed, existence of a horizon magnifies quantum effects in presence of a black hole, revealing a gamut of exotic features, the most famous being its thermal attributes \cite{bhthermod-revs}. This {\it gravity $\leftrightarrow$ quantum $\leftrightarrow$ thermodynamics} connection has been gaining increasing attention in recent years due to it's potential relevance for our understanding of gravity, and perhaps spacetime itself, at a fundamental level. In particular, the fact that black holes have thermodynamic properties has time and again motivated the intriguing question: {\it is there a deeper clue hidden in the connection between gravitational dynamics and thermodynamics}? One is naturally lead to suspect whether the \textit{dynamics} of gravity itself has a built in thermodynamic structure. Moreover, since these thermal attributes are essentially quantum mechanical in origin, they are particularly relevant in the context of quantum gravity.

This article zooms into two very specific questions in this context:
\begin{description}[Type 1]
\item[${\bm \star}$]{What information about {\textbf{spacetime curvature}} can be obtained from {\textbf{thermal properties}} of local {\textbf{acceleration}} horizons?}

The first part of this article largely focuses on exploring the spacetime geometry in the vicinity of an accelerated observer in the hope that it will clarify the connection between gravity and thermodynamics. Specifically, I will try to clarify the relation between {\it area variation} and Einstein tensor
$$
T \, \underset{?}{\delta} \mathcal A \underset{?}{\longleftrightarrow} {\bm G}(\bm u_{\rm R}, \bm u_{\rm R})
$$
by trying to answer each of the $``?"$s in the above expression. Such a relation has been known and discussed in various forms in several works; I hope the analysis given here would help clarify it's mathematical origin. 

\item[${\bm \star}$]{What are the effects of {\textbf{spacetime curvature}} on {\textbf{thermal properties}} of a {\textbf{freely falling}} quantum system?}

I then present some speculations on geodesic free fall of a thermal system across the local horizon of such an observer. Motivated by results from a couple of examples, I conjecture that the partition function and thermal entropy of {\it any} quantum system at high temperatures acquire a system independent correction term characterised by the dimensional quantity 
$$\Delta = {\bm R({\bm u}_{\rm ff}, {\bm u}_{\rm ff})} \l(\hbar c/kT\r)^2$$
which contributes an (entropy)/(degree of freedom) of 
\begin{eqnarray}
s_{\mysymb} = ({\rm const.}) \; {\bm R({\bm u}_{\rm ff}, {\bm u}_{\rm ff})} \l( {\hbar c}/{kT} \r)^2
\label{eq:euler2}
\end{eqnarray}
\end{description}
Each of the highlighted words in the above two questions carries physical significance of it's own; note that the only difference between the two is that of accelerated vs. inertial motions. 

I feel a deeper analysis of these questions is important for any research that uses the connection between gravity and thermodynamics as a starting point. One example would be the assertion that there is more to gravitational dynamics than the Einstein-Hilbert action -- something that has been the focus of Paddy's research over the past decade. This article presents some of the lesser known results, both mathematical and physical, that could be important in a better understanding of such an assertion.

\blfootnote{
\textit{Notation:} The metric signature is $(-,+,+,+)$; latin indices go from $0$ to $3$, greek indices from $1$ to $3$. 
}
\section{Gravity $\Longleftrightarrow$ Accelerated frames $\Longleftrightarrow$ Thermodynamics}
\label{sec:2}
Once we realise that certain surfaces - {\it horizons} - must be attributed entropy by certain class of observers, there is a tantalizing possibility of \textit{imposing} the first law of thermodynamics and then see how it constrains the curvature of background spacetime. It is possible that such a study can also yield information about Einstein field equations themselves. 

One physically relevant set-up for addressing this issue was given by Jacobson, based on usage of local acceleration (Rindler) horizons and the Raychaudhuri equation \cite{jacobson-eos}. Subsequently, several generalisations as well as variations of this work have appeared in the literature. However, not all of them share the same physical and/or technical assumptions. Several such issues have been discussed in \cite{thermod-diffs1, thermod-diffs2}. At some level, some of these issues do become important 
when one actually probes the structure of gravitational field equations and look for some connection with some thermodynamic law, a route that was first taken by Padmanabhan and has subsequently been studied in much more generality \cite{TP-thermod}. The relevant structure of field equations (or technically, the Einstein tensor) is of obvious relevance to any argument(s) which attempt to reverse the logic and derive field equations from thermodynamic considerations. A clarification of some of the subtle mathematical issues involved in this program was presented in \cite{thermod-diffs1, thermod-diffs2}.

In this note, we present a calculation directly in the local frame of the accelerated observer, described by Fermi coordinates, that illuminates several of the points concerning the structure of Einstein tensor and it's role in interpreting the field equations in thermodynamic terms. That is, we investigate the local spacetime structure by employing, as probes, accelerated trajectories and using the fact that such trajectories will have a local horizon attributed with the standard Unruh temperature \cite{DK-thesis}.

This analysis is therefore complimentary, though not completely identical, to the one presented in \cite{thermod-diffs1}. It has the advantage that it very clearly demonstrates how various curvature components combine, in a rather specific manner, \textit{through the transverse horizon area}, to reproduce the relevant (time-time) component of the Einstein tensor. {\it This mathematical fact lies at the heart of any attempt to relate gravitational dynamics with horizon thermodynamics.} From a broader perspective, the calculations presented here also re-emphasize the points made earlier in \cite{thermod-diffs1} concerning the difference(s), both technical and conceptual, between results arising from exploring thermodynamic structure of field equations and attempts to derive the field equations from a specific thermodynamic statement (see \cite{thermod-diffs2} for a more recent comprehensive discussion of these issues); in particular, the former is \textit{not} implied by the latter 

The key point clarified and highlighted by the analysis presented here is {\it how} and {\it why} the ``time-time" component of the field equations (equivalently, of the Einstein tensor) acquires a direct thermodynamic relevance
\footnote{It is always this component that has appeared in works based on \cite{TP-thermod}; it's significance in related contexts is also becoming evident in some recent works (compare, for example, \cite{jacobson-ee} and \cite{jacobson-eos}).}.
This forms the main focus of this paper. Appendix \ref{app:toy-model} gives a simple toy model for quantifying the shift in location of the horizon as some form of energy crosses it. Appendix \ref{app:surface-term} presents the (leading order) form of the surface term in the Einstein-Hilbert action in the local coordinates of the accelerated observer in curved spacetime. This term is known to yield horizon entropy in flat spacetime, and hence one expects the curvature corrections to the same as also having direct thermodynamic significance.

\subsection{Virtual displacements of local ``Rindler" horizons and Einstein tensor}
\label{subsec:2p1}
It is well known that accelerated observers in \textit{flat spacetime} perceive the Minkowski vaccuum as a thermal bath; therefore, it is natural to start looking for the connection between gravity and thermodynamics by considering accelerated observers in \textit{curved spacetime}. Consider, then, such an observer (or, more precisely, an accelerated timelike trajectory), and construct a locally inertial system of coordinates based on his/her trajectory. Such a coordinate system can be obtained by Fermi-Walker transporting the observer's orthonormal tetrad along the trajectory; the coordinate system so obtained is called the Fermi coordinate (FNC) system \cite{fnc-refs}. In FNC's $(\tau, x^\mu)$, the metric acquires the following form:
\begin{eqnarray}
g_{0 0} &=& - \l[ \l( 1 + a_{\mu} x^{\mu} \r)^2 + R_{0 \mu 0 \nu} x^{\mu}x^{\nu} \r] + O(x^3)
\nn \\
g_{0 \mu} &=& - \frac{2}{3} R_{0 \rho \mu \sigma} x^{\rho} x^{\sigma} + O(x^3)
\nn \\
g_{\mu \nu} &=& \delta_{\mu \nu} - \frac{1}{3} R_{\mu \rho \nu \sigma} x^{\rho} x^{\sigma} + O(x^3)
\end{eqnarray}
where $a^\mu$ and $R_{abcd}$ are all functions of $\tau$. We shall assume that we are working in a sufficiently small region of space, and focussing on sufficiently small time scales, so as to ignore the time dependence of metric components. Technically, this is equivalent to assuming the existence of a static timelike Killing vector in the region of interest. For definiteness, we will take our observer to be moving along the $x^3$ direction; 
i.e., $a^{\mu} = a \delta^{\mu}_3$, where $a$ is the norm of the acceleration. This can be always be done by an 
appropriate choice of the basis vectors: Given the four velocity ${\bm u}$ of the observer, one chooses 
the basis vectors such that, ${\bm e}_{0} = {\bm u}$ and ${\bm e}_{3} = (1/a)\;{\bm a}$, so 
that ${\bm e}_{0} \cdot {\bm e}_{3} = 0$. These basis vectors are then Fermi-Walker transported along the 
trajectory, $\nabla_{\bm u} {\bm e}_k = \l( \bm{ u \otimes a - a \otimes u } \r) \cdot {\bm e}_k$ for $k=0 \; .. \;3$. 
The remaining two basis vectors can be suitably chosen to be orthogonal to ${\bm e}_{0}$ and ${\bm e}_{3}$. In such a 
coordinate system, the only non-zero component of the acceleration 4-vector is ${\bm a}_{3} = {\bm a} \cdot {\bm e}_{3} = a$. 

The FNC system (to quadratic order) describes the geometry in the neighbourhood of an accelerated observer to a very good accuracy for length 
scales $x \ll {\mathrm{min}} \; \{\; |a|^{-1}, R^{-1/2}, | R / \partial R | \;\} $. We must now choose the acceleration such that the length scale $|a|^{-1}$ associated with it is much smaller than the curvature dependent terms above; then the presence of the horizon along with it's thermodynamic attributes will not be affected by curvature. Specifically, the horizon for such an observer is then located at $x^3 = - a^{-1} = z_0$, say. 

The transverse area of the horizon 2-surface, $x^3 = z_0$, is given by
\begin{eqnarray}
\sigma_{\chf{AB}} = \delta_{\chf{AB}} - \frac{1}{3} R_{\chf{A} \mu \chf{B} \nu} x^{\mu} x^{\nu}
\end{eqnarray}
so that
\begin{eqnarray}
\sqrt{\sigma} &=& 1 - \frac{1}{6} R_{\chf{A} \mu \chf{A} \nu} x^{\mu} x^{\nu}
\\
&=& 1 - \frac{1}{6} \l[ R^{\chf{A} 3}_{\; \chf{A} 3} z_0^2 + 2 R^{\chf{A} 3}_{\; \chf{A} \chf{B}}\, z_0\, y^{\chf{B}} + R^{\chf{AC}}_{\; \chf{BD}} x^{\chf{C}} x^{\chf{D}} \r]
\end{eqnarray}
where we have used the Euclidean metric to raise the spatial indices on the curvature tensor, since the curvature tensor is evaluated on the observers' worldline.

We now wish to consider some physical process which produces a virtual displacement of the horizon normal to itself (that is, along the $x^3$ direction), and see how the horizon area changes in such a process (see Appendix \ref{app:toy-model} for a toy model of such a process). The displacement is to be considered as a {\it virtual displacement}; we do \textit{not} have in mind an observer with a different acceleration, which would require us to construct a different coordinate system based on the new worldline. Rather, the idea here is to see whether the differential equation governing the virtual displacement of the horizon has any physical meaning. We shall therefore consider the parametric dependence of the transverse area on $z_0 = -a^{-1}$, $\mathcal{A}(x, y; z_0)$, and then consider the variation, 
\begin{eqnarray}
z_0 &\rightarrow& z_0 + \epsilon
\nn \\
\epsilon &=& \Delta z_0
\end{eqnarray}
Since we generally associate an entropy $S \propto \mathcal{A}$ with a horizon surface, and a local Unruh temperature $T_U = a / (2 \pi)$ with an accelerated observer, we expect the resultant equation for involving $\delta \mathcal{A}$ to have a thermodynamic interpretation.

We have,
\begin{eqnarray}
\delta_{z_0} \sqrt{\sigma} &=& \sqrt{\sigma} |_{z_0+\epsilon}  - \sqrt{\sigma} |_{z_0}
\nn \\
&=& - \frac{1}{3} \l[ R^{\chf{A} 3}_{\; \chf{A} 3} z_0 + R^{\chf{A} 3}_{\; \chf{AB}} \; x^{\chf{B}} \r] \epsilon + O(\epsilon^2)
\label{eq:link1}
\end{eqnarray}
where the subscript $\epsilon$ reminds us that we are dealing with a very specific variation. Now concentrate on a small patch of the horizon surface. It is then not unreasonable to assume that, upon integration over the transverse coordinates, the second term in the square brackets averages out to zero. (If it doesn't, we would have a preferred direction in the transverse horizon surface.) We shall nevertheless come back to this point later. 

The change in area of this surface under the virtual displacement of the horizon is therefore given by 
\begin{eqnarray}
\delta_{z_0} \mathcal{A} &=& \int \limits_{\mathcal H} \l( \delta_{z_0} \sqrt{\sigma} \r) \DM^2 x_{\perp}
\end{eqnarray}
where $\DM^2 x_{\perp} = \DM x^1 \DM x^2$, and $\mathcal{H}$ represents integration over the horizon surface $\tau=$\, constant, $x^3=z_0$. With $T_U = -1/(2 \pi z_0)$ for $a > 0$, we have
\begin{eqnarray}
T_U \; \delta_{z_0} \l( \frac{1}{4} \mathcal{A} \r) = \frac{\eta}{8 \pi}  \int \limits_{x^3=z_0} R^{3 \chf{A}}_{\; 3 \chf{A}} \; \epsilon \; \DM^2 x_{\perp}
\end{eqnarray}
where $\eta = 1/3$. If one assumes the standard Bekenstein-Hawking expression for entropy of a horizon, $S=\mathcal{A}/4$, then the left hand side above is just $T_U \; \delta S$. To simplify the right hand side, we use the general decomposition of the Einstein tensor in terms of components of the curvature tensor. This is given by
\begin{eqnarray}
G^0_{\; 0} = - \l( R^{1 2}_{\; 1 2} + R^{1 3}_{\; 1 3} + R^{2 3}_{\; 2 3} \r)
\end{eqnarray}
which implies
\begin{eqnarray}
R^{3 \chf{A}}_{\; 3 \chf{A}} = - G^0_{\; 0} - R^{12}_{\; 12}
\end{eqnarray}
Then, we obtain
\begin{eqnarray}
T_U\, \delta_{z_0} S = - \eta \l[\,  \int \limits_{\mathcal H} \frac{1}{8 \pi}\, G^0_{\; 0}\, \epsilon \, \DM^2 x_\perp + \int \limits_{\mathcal H} \frac{1}{16 \pi} \, R_{\parallel} \, \epsilon \, \DM^2 x_\perp \r]
\nn \\
\label{eq:tds-1}
\end{eqnarray}
where $R_{\parallel} = R^{\chf{AB}}_{\; \chf{AB}}$ is the Ricci scalar of the in-horizon 2-surface. This is essentially the relation we wanted to establish. Note that, to the relevant order, one can consider $\DM^2 x_\perp$ on the RHS of Eq.\;(\ref{eq:tds-1}) as the covariant measure of horizon area; since it is already multiplied by curvature components, any curvature corrections to volume would be higher order. The appearance of $G^0_{\; 0}$ is already quite suggestive, and, in what follows, we argue that the second term also has a nice interpretation which qualifies it as a suitable thermodynamic variable.

\subsubsection{Horizon energy}
We now provide a suitable interpretation for the second integral in Eq.\;(\ref{eq:tds-1}). We begin by noting that the Euler character $\chi$ of a two dimensional manifold $\mathcal{M}_2$, in this case the horizon 2-surface, is given by
\begin{eqnarray}
\chi \left( \mathcal{M}_2 \right) = \frac{1}{4 \pi} \int_{\mathcal{M}_2} ~ R ~ \mathrm{d[vol]} \; + \; \mathrm{boundary~terms}
\end{eqnarray}
The second term in the square brackets in Eq.\;(\ref{eq:tds-1}) therefore just $( \chi / 4) \epsilon $ plus the boundary term, which involves the integral of extrinsic curvature over the boundary of the region of integration. Keeping all these points in mind, we will simply call the second integral in Eq.\;(\ref{eq:tds-1}) as $\hat{\chi}$. We want to interpret this term as change in ``energy" associated with the horizon, say $E_g$. This expression is already well known in the context of quasilocal energy in spherically symmetric spacetimes. Let us consider the case of a general, spherically symmetric black hole. This term then essentially involves the curvature component, $R^{\theta \phi}_{~~\theta \phi} = (1/r^2)\;(1 - g^{rr})$ in standard coordinates. On the horizon $r=a$, $g^{rr}$ vanishes and we obtain, after multiplying with the appropriate transverse area element, $E_g=a/2$, which is a very common expression for energy (called as the Misner-Sharp energy). In Appendix \ref{app:toy-model}, we give a toy model which further analyses this expression for (change in) energy in terms of horizon shift.  

\subsection{Final Result and Discussion}
\label{subsec:3}
If one employs the Einstein field equations $G^0_{\; 0} = 8 \pi T^0_{\; 0}$ at this stage (where $T^0_0 = - \rho_m$ is the local energy density of matter in the instantaneous rest frame of the observer), Eq.\;(\ref{eq:tds-1}) becomes 
\begin{eqnarray}
T_U \; \delta_{\epsilon} S &=& - \eta \l[ \l( \int \limits_{\mathcal H} T^0_{\; 0} \DM^2 x_\perp \r) \; \Delta z_0 + \frac{\hat{\chi}}{2} \l( \frac{\Delta z_0}{2} \r) \r]
\nn \\
\label{eq:tds-2}
\end{eqnarray}
where the second equality is based on discussion of the previous section. This is our main result, which forms the basis for thermodynamic interpretation of field equations. For example, for static spacetimes, the above relation (apart from the factor $(-\eta)$), can be shown to have the form \cite{dk-tp}
\begin{eqnarray}
T \delta S &=& \bar{F} \Delta z_0  + \DM E_g 
\end{eqnarray}
where $\bar{F}$ is the average normal force on the horizon surface (see Eq. (21) of \cite{dk-tp}).

It is gratifying to see that the final expression is in a form which can be readily expressed in any arbitrary coordinate system; it only depends upon the foliation provided by the accelerated observer. Given $\bm u$ and $\bm a$, one can identify the spacelike two-surface acting as local horizon, and the corresponding ``heat" flow then depends only on $R_{\parallel}$ -- the curvature scalar of the two surface, and $G^0_{\; 0} = - {\bm G({\bm u}, {\bm u})}$ -- which is the projection of Einstein tensor along the observer's time axis. This is as far as one can get trying to explore the connection between intrinsic properties of local Rindler horizons and gravitational dynamics. 

To evaluate how rigorous the analysis is, let us take stock of the assumptions that have gone into the derivation: 

(i) The acceleration length scale be small compared to any of the curvature length scales: This assumption is natural, given that the whole idea is to use solely the acceleration of probe observes and find constraints on background geometry - a natural physical setup to formulate the problem, sanctioned by the equivalence principle. 

(ii) The assumption mentioned just below Eq.~(\ref{eq:link1}): This requires some consideration, since it is possible that the term contributes in a sensible manner to some form of energy associated with some geometric charactersitic of the horizon. It might introduce additional stresses in the first law (similar to those resulting from, say, angular momentum of the horizon patch).

The only technical issue is that, this derivation, while as general as it can be, yields an extra factor of $-\eta=-1/3$ whose interpretation is unclear. Otherwise, the analysis very clearly indicates how the gravitational dynamics of a curved spacetime is intimately related to the property of spacelike two surfaces which can act as horizons for certain observers. In fact, as we have shown above, the gravitational dynamics (or more specifically, the Einstein tensor) is completely encoded in the transverse geometry of the spacelike two surface and it's normal variation, to which a physical interpretation can be attached via thermodynamic quantities such as entropy and temperature associated with the horizon \cite{makela}.

It is indeed curious that we can make any statement at all concerning the dynamics of gravity by (i) assuming a curved Lorentzian spacetime with it's associated light-cones, and (ii) a temperature associated with local Rindler horizons. Curiously, the curvature tensor makes it's appearance through the thermodynamic variables $T_U$ and $\delta S$, rather than the conventional {\it tidal} terms. The reason is that, apart from the laboratory length scale, we now also have the length scale $c^2/a$ associated with acceleration. The presence of a causal horizon at $z_0=-1/a$ and it's associated thermodynamics is what brings in the curvature information through {\it quantum} effects. The non-trivial part being that it brings in just the right combination of curvature components to facilitate making a general statement about the dynamics of gravity. 
\begin{svgraybox}
\textit{More succinctly, just as \textbf{accelerated frames} played a key role in arriving at a \textbf{kinematic} description of gravity in terms of spacetime geometry, \textbf{virtual displacements of acceleration horizons}, through associated quantum effects, might play a key role in understanding better the \textbf{dynamics} of gravity.
}
\end{svgraybox}

We hope we have provided one of the most straightforward demonstrations of the above ideas. In the next section, we focus attention to the fate of actual thermal systems in free fall in a curved spacetime, and conjecture on a universal tiny contribution to thermal entropy of {\it any} system at sufficiently high temperatures.

\section{Thermal entropy and spacetime curvature}
\label{sec:3}
Most of the work investigating the connection between gravity and thermodynamics so far has focussed on accelerated observers and the thermal effects associated with their horizons. However, all such analyses invariably are based on a scenario in which a system disappears across the acceleration horizon, carrying 
entropy with it that is {\it lost} to the accelerated observer. Indeed, it is this question (posed in the context of black holes) of Wheeler's that had prompted Bekenstein to investigate the physical basis behind the laws of black hole mechanics \cite{bekenstein1}. Unfortunately, however, not much attention has been given to the statistical mechanics of quantum 
systems that are (in some suitably defined sense) in a free fall in a curved spacetime. 

It is precisely this problem that motivated the analysis in \cite{dk-idealgas}, two of the key points of interest being: (i) to study the interplay between quantum and thermal fluctuations in presence of spacetime curvature, (ii) to analyse whether this can have any implications for understanding the physical nature of black hole entropy. Here, I will focus only on point (i). Since a general analysis of statistical mechanics of a quantum system at finite temperature, in a curved spacetime, is understandably difficult, the above questions were addressed in \cite{dk-idealgas} in the context of a very simple system: non-relativistic ideal gas enclosed in a box whose (geometric center) follows a geodesic $\bm u_{\rm ff}$. The choice of central geodesic only serves to define the reference measure for energy of the system; in particular, the Hamiltonian of the particles is defined as 
\begin{eqnarray}
H = - c\, p_{\widehat 0}
\end{eqnarray}
where ${\widehat 0}$ is the time coordinate defined by $\bm u_{\rm ff}$.
More discussion on the physical significance of this Hamiltonian, as well as it's detailed form, can be found in \cite{dk-idealgas}. In the non-relativistic limit, one obtains
\begin{eqnarray}
H - m c^2 \approx \frac{p^2}{2 m} + \frac{1}{2} m c^2 R_{{\widehat 0} \mu {\widehat 0} \nu} y^{\mu} y^{\nu} 
\end{eqnarray}
I will also ignore the time dependence carried by curvature components, a reasonable assumption if the time scale $\mathcal{R}/\mathcal{\dot R}$ is much larger compared to typical time scale associated with the system; one expects this to be the case at high temperature (see below). 

One can now use the above Hamiltonian to evaluate the energy eigenvalues $E\l[ \{n\} \r]$ of constituent particles of a system (where $\{n\}$ represents the set of all relevant quantum numbers) at temperature $kT=1/\beta$, and therefore the partition function
\begin{eqnarray}
Z(\beta) &=& \sum \limits_{\{n_i\}} 
e^{ - \beta E\l[\{n_i\}\r] }
\label{eq:fullz}
\end{eqnarray}
The one-particle energy eigenvalues have the form
\begin{eqnarray}
E_1\l[ \{n_i\} \r] &=& E_{1,0}\l[ \{n_i\} \r]
+
\Delta E_1\l[ \{n_i\} \r]
\;\;\;\;\;\;\;\;\;\;\;\;
\label{eq:pert-eev-gen}
\end{eqnarray}
where the first term on RHS represents unperturbed energy eigenvalues, while the second term is the perturbation cause by spacetime curvature. The corresponding partition has the form
The one-particle energy eigenvalues have the form
\begin{eqnarray}
Z(\beta) = Z_0(\beta) + \Delta Z\l(\beta, R_{abcd}, \{\varkappa\} \r)
\label{eq:pert-pf}
\end{eqnarray}
where again, the first term on RHS is the flat spacetime expression and the second term the first order modification due to spacetime curvature; $\{\varkappa\}$ symbolically denotes all the system specific parameters, such as mass, physical dimensions etc. 

Our key conjecture would be that:
\begin{eqnarray}
\lim \limits_{{\rm large} \; T} \Delta Z\l(\beta, R_{abcd}, \{\varkappa\} \r) 
\end{eqnarray}
generically contains a term independent of $\{\varkappa\}$, and has the form 
\begin{eqnarray}
\lim \limits_{{\rm large} \; T} \Delta Z\l(\beta, R_{abcd}, \{\varkappa\} \r) &=& ({\rm const.}) R_{00} \Lambda^2 +
\mathrm{ \{\varkappa\}~dependent~terms}
\end{eqnarray}
Once again, such a result would imply that there is a universal, inherent, Ricci contribution to thermodynamic quantities of a system in curved spacetime, which, for obvious reasons, can have great implications when a system at finite temperature disappears across a causal horizon of some observer. 

\subparagraph{Box of ideal gas} The calculation mentioned above was carried out in \cite{dk-idealgas} for a box of ideal gas freely falling in a curved spacetime, and it was shown that, in the regime where the analysis was valid ($\lambda^3/V \ll 1$ where $\lambda = h/\sqrt{2 \pi m k T}$ is the thermal de Broglie wavelength),
%
the partition function has the form:
\begin{eqnarray}
\ln \l( \frac{Z}{Z_{\mathrm F}}
\r)
&=&
-\frac{1}{4} N R_{{\widehat 0 \widehat 0}} \Lambda^2 + {\rm terms~depending~on~curvature~\&~box~details}
\label{eq:partitionfn}
\end{eqnarray}
where $\ln Z_{\mathrm{F}}=\ln (V^N \lambda^{-3N}/N!)$ is the flat space expression, and $\Lambda=\beta \hbar c$ -- a length scale independent of box dimensions $L_i$ and mass $m$. 

\noindent We can now obtain corrections to various thermodynamic quantities: $U_{\mathrm{corr}} = U-U_{\mathrm{F}}$ and $S_{\mathrm{corr}} = S-S_{\mathrm{F}} $, where $U_{\mathrm{F}}=3N/2\beta$ and $S_{\mathrm{F}} = 3N/2 + N \ln \l(e V/N \lambda^3\r)$ are standard flat space expressions. Using standard definitions $U=-\partial_{\beta} \ln Z$ and $S=\ln Z + \beta U$ to evaluate $U_{\mathrm{corr}}$, $S_{\mathrm{corr}}$ and heat capacity at constant volume, $C_V=-\beta^2 \partial_{\beta} U = 3N/2 + C_{V \mathrm{corr} }$, we obtain
\begin{equation*}
\begin{rcases}
{2 S_{\mathrm{corr}}}/{N} &= 
\underbrace{+ (1/2) R_{{\widehat 0 \widehat 0}} \Lambda^2}_{2 s_{\mysymb}/N}  \hspace{.25cm}
\nn \\
{\beta U_{\mathrm{corr}}}/{N} &= 
\underbrace{+ (1/2)  R_{{\widehat 0 \widehat 0}} \Lambda^2 }_{\beta u_{\mysymb}/N}
\nn \\
{C_{V \mathrm{corr} }}/{N} &= \underbrace{- (1/2) R_{{\widehat 0 \widehat 0}} \Lambda^2}_{c_{\mysymb}/N}  
\nn \\
\end{rcases}
\text{+ system dependent terms}
 \label{eq:final-expr}
\end{equation*} 
%

Before proceeding, it is worth pausing to check whether the approximations made to arrive at the above result(s) are not mutually inconsistent. This is an important issue, and is discussed at length in  Appendix \ref{app:idealgas-approxs}.

\subparagraph{Simple harmonic oscillator} One can also do the above analysis for a bunch of simple harmonic oscillators with frequency $\omega$ \cite{dk-unpublished} (which might be physically more relevant system for obvious reasons); in this case, one obtains, in the high temperature limit $\beta \hbar \omega \ll 1$
\begin{equation*}
\begin{rcases}
{2 S_{\mathrm{corr}}}/{N} &= 
\underbrace{+ (1/12) R_{{\widehat 0 \widehat 0}} \Lambda^2}_{2 s_{\mysymb}/N}  \hspace{.25cm}
\nn \\
{\beta U_{\mathrm{corr}}}/{N} &= 
\underbrace{+ (1/12)  R_{{\widehat 0 \widehat 0}} \Lambda^2 }_{\beta u_{\mysymb}/N}
\nn \\
{C_{V \mathrm{corr} }}/{N} &= \underbrace{- (1/12) R_{{\widehat 0 \widehat 0}} \Lambda^2}_{c_{\mysymb}/N}  
\nn \\
\end{rcases}
\text{+ system dependent terms}
\end{equation*} 

The system dependent terms not indicated above can be found in \cite{dk-idealgas, dk-unpublished}, and they essentially involve terms that depend on $m, L_i$ in the case of ideal gas, and the frequency $\omega$ in the case of harmonic oscillator.

I now wish to highlight some of the key features of this result:
 
\begin{itemize} 
\item[$\star$] Perhaps the most important point to be noted is the following: For the example of ideal gas, the curvature dependent correction term $\Delta E_1\l[ \{n_i\} \r]$ in Eq.~(\ref{eq:pert-eev-gen}) turns out to be independent of $\hbar$! (The full expression is given in \cite{dk-idealgas}.) The quantum{\it ness} of this term is solely due to the $\{n_i\}$ dependence of $\Delta E_1\l[ \{n_i\} \r]$, and manifests itself in the final expressions since $\Lambda \propto \hbar$. This is a nice demonstration of how the interplay between quantum and thermal fluctuations induced by a background spacetime curvature can be non-trivial.

\item[$\star$] Further, $s_{\mysymb}$ and $u_{\mysymb}$ satisfy the relation: $s_{\mysymb} = ({1}/{2}) \beta u_{\mysymb}$
with 
$s_{\mysymb}$ as mentioned in the Introduction (Eq.~(\ref{eq:euler2})). This is a Euler relation of homogeneity two, well known from black hole thermodynamics; in particular, black hole horizons have temperature $\beta^{-1}_H$, entropy $S_{\rm bh}$ and (Komar) energy $U_{\rm bh}$ which also satisfy $S_{\rm bh}=(1/2) \beta_H U_{\rm bh}$. 
%
Relevance of such Euler relation and area scaling of entropy for self-gravitating systems has already been emphasized in \cite{areascaling}. This relation also plays an important role in the {\it emergent gravity} paradigm, leading to an {\it equipartition law} for microscopic degrees of freedom associated with spacetime horizons \cite{paddy-equip}. 

\item[$\star$] The $\mysymb$ contribution to specific heat is {\it negative} if the condition ($R_{{\widehat 0} {\widehat 0}} \geq 0$) holds. (This condition is, of course, tied to the strong-energy condition if Einstein equations are assumed.) Also, $c_{\mysymb}=-\beta u_{\mysymb}=-2s_{\mysymb}$, which are again the same as the relations satisfied by a Schwarzschild black hole.

\item[$\star$] The appearance of the length scale $\Lambda=\hbar c/kT$ in the non-relativistic limit is curious, and it would be interesting to understand the physical meaning of this length scale at the basic level \cite{haggard-rovelli}. 
\end{itemize}

\subsection{Speculation}
All the above points are extremely suggestive as far as the role of Ricci correction to thermodynamic properties of arbitrary systems is concerned. In fact, a similar analysis can be done for a harmonic oscillator, and it can again be shown that at sufficiently high temperatures, thermodynamic quantities acquire specific correction terms which are independent of the frequency $\omega$ of the oscillator \cite{dk-unpublished}.

In particular, based on these, one can speculate the following:

\begin{svgraybox}
\textit{Entropy of a system at temperature $T$ generically acquires a system independent contribution in a curved spacetime characterized by the dimensionless quantity  
\begin{eqnarray}
\Delta = {\bm R({\bm u}_{\rm ff}, {\bm u}_{\rm ff})} \l(\hbar c/kT\r)^2
\end{eqnarray}
at sufficiently large temperatures $T$.
}
\end{svgraybox}

This should motivate further study of thermal systems in curved spacetime along the suggested route. Note that the important features associated with the Ricci corrections would not appear if: (i) one studies the 
problem using {\it classical} statistical mechanics -- since then the modification of energy eigenvalues, $1/n^2$ in present case, is missed), or (ii) assume a priori that {\it finite size corrections} would necessarily depend only on area, perimeter etc -- the Ricci term here in fact does not involve dimensions of the box at all! For the result to have any deeper significance, it's main {\it qualitative} aspects (in the high temperature limit) must, of course, survive further generalizations (relativistic gas, different statistics etc.), insofar as the {\it form} of the Ricci term is concerned. Some preliminary calculations do seem to point to this \cite{dk-unpublished2}.

\subsection{Possible Implications}
So, what could be possible implications of this? 

Well, for one, if there does exist a universal term in entropy of systems at high temperature that depends on the Ricci tensor, it might lead to some insight into understanding the interplay between thermal and quantum fluctuations in a curved spacetime. 

Second, such analyses can add interesting (and non-trivial) physics to arguments given by Bekenstein in his original paper \cite{bekenstein1} in support of the so called generalised second law (GSL). Most of Bekenstein's supporting analysis used expressions for thermal energy and entropy of various systems in flat spacetime, added to the (minimum) change in black hole entropy when such a system falls across the black hole horizon. It is, of course, important and interesting to know how curvature corrections to thermodynamic attributes of a system affect this analysis. 

More specifically, one would like to know {\it quantitatively} how and where the $\mysymb$ term appears in the proposed GSL:
$$
\Delta S_{\rm BH} + \Delta S_{\rm ext} > 0
$$
where $\Delta S_{\rm BH}$ represents the change in entropy of the black hole, and $ \Delta S_{\rm ext} =  \Delta S_{\rm ext} \l(\beta, R_{abcd}, \{\varkappa\} \r)$ is the change in common entropy in the region exterior to the black hole. This is work under progress.

And lastly, appearance of such a term can be of direct relevance for understanding of thermodynamical aspects of gravitational dynamics.

\section{Concluding remarks}
\begin{quotation}
Now, here, you see, it takes all the running you can do, to keep in the same place. If you want to get somewhere else, you must run at least twice as fast as that!

\hfill - Lewis Carroll, {\it Through the Looking Glass} 
\end{quotation}

        Since all the relevant comments and remarks have been given in the respective sections, I conclude it with the following pictorial summary of the theme of this article:
        \begin{center}
        \begin{figure*}[h!]%
            \subfloat{{\includegraphics[width=.85\textwidth]{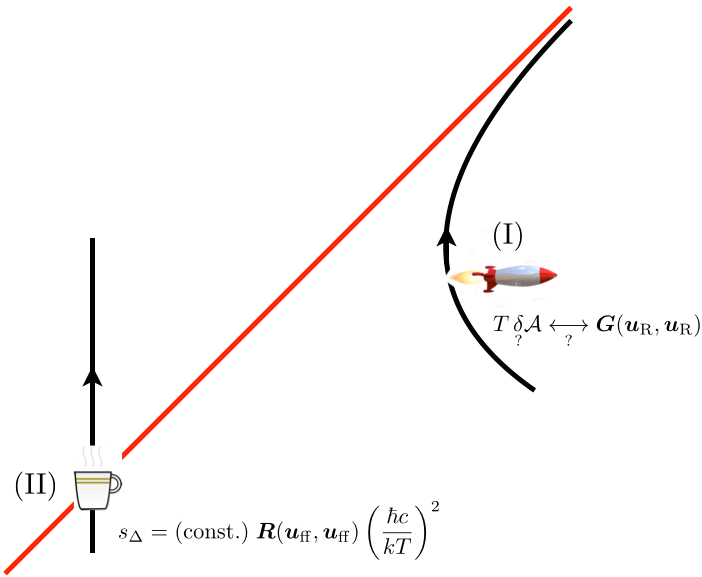} }}%
            \caption{{A combined description of a freely falling thermal system {\it and} the thermodynamics associated with an accelerated observer can yield physically useful insights into interplay between quantum mechanics, thermodynamics, and spacetime curvature}.}%
            \label{fig:summary}%
        \end{figure*}
        \end{center}
\newpage Study of thermodynamic aspects of gravity, which derives it's motivation from semi-classical results such as the Hawking and Unruh effects, does provide an elegant route to make some deep observations concerning the nature of gravity. However, such considerations, by themselves, might turn out to 
be {\it too elegant} to be of any use, unless coupled with a deeper study of the structure of statistical mechanics in a curved spacetime. 

\begin{acknowledgement}
I thank Paddy for many discussions and comments on these and related topics over a course of almost ten years. The support of Department of Science and Technology (DST), India, through it's INSPIRE Faculty Award, is gratefully acknowledged.
\end{acknowledgement}
%
%
%
\section*{Appendix}
\section{A simple toy model for a point mass disappearing across the horizon} \label{app:toy-model}
\addcontentsline{toc}{section}{Appendix}
In this section, let us try to model the loss of energy across a local Rindler horizon via a particle of mass $m$ that disappears across the horizon. We will present the analysis for the case when the background spacetime is flat in the limit $m \to 0$; that is, $m$ is the only source of curvature. As we shall comment in the end, this suffices as long as one is working to first order in background curvature.
\begin{figure*}%
    \centering
    \subfloat[Inertial coordinates]{{\includegraphics[width=0.4\textwidth]{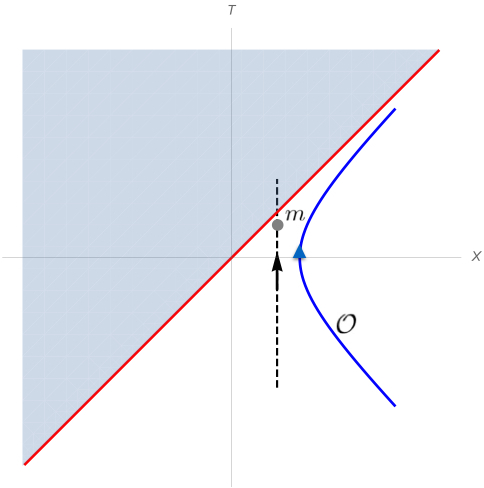} }}%
    \hspace{2cm}
    \subfloat[Rindler coordinates]{{\includegraphics[width=0.4\textwidth]{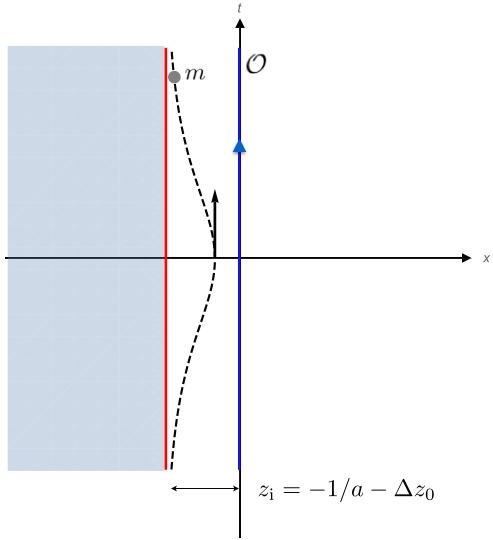} }}%
    \caption{A particle of mass $m$ crossing a local Rindler horizon of $\mathcal O$; Inertial and Rindler perspectives.}%
    \label{fig:rindler-geod}%
\end{figure*}

\begin{figure*}%
\sidecaption[b]
    \centering
    \subfloat{{\includegraphics[width=0.6\textwidth]{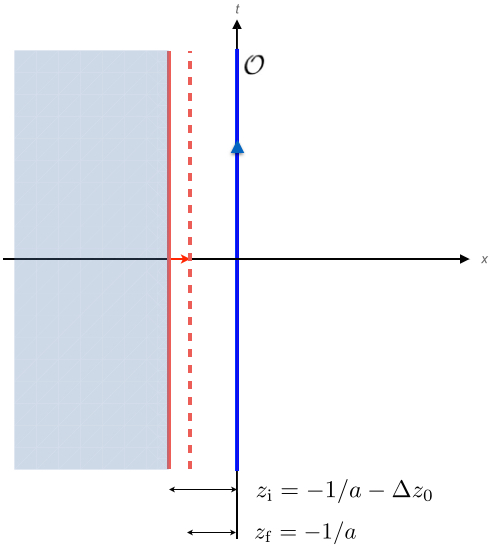} }}%
    \caption{The shift of horizon due to ``loss" of mass $m$ (see text for details).}%
    \label{fig:hor-shift}%
\end{figure*}

A complete dynamical description of this process is expected to be complicated, but since we are only interested in shift of the horizon as the particle disappears across it, we propose the following simplified scenario: we consider the particle when it is on the verge of crossing the horizon, that is, at $x^{\mu} \approx -(1/a) \delta^\mu_3 - 0, t=t_0$, and compare this with a situation when it is no longer in the causal domain of the accelerated observer. The shift in horizon position can then be evaluated by using an orthonormal tetrad for this particle that maps smoothly to that of the accelerated observer (in an asymptotic sense, see below). This is most easily done by using the Schwarzschild metric in isotropic coordinates $y^\mu$ for the particle; the metric near $m$ in these coordinates is given by

\begin{eqnarray}
\DM s^2 &=& - \l( \frac{1 - \frac{m}{2r}}{1 + \frac{m}{2r}} \r)^2 \DM t^2 + \l( 1 + \frac{m}{2r} \r)^4 \DM l_{\mathrm{flat}}^2
\end{eqnarray}
where
\begin{eqnarray}
\DM l_{\mathrm{flat}}^2 &=& \delta_{\mu \nu} \DM y^\mu \DM y^\nu  
\\
r &=& \sqrt{\delta_{\mu \nu} y^\mu y^\nu}
\end{eqnarray}
and we shall be interested in $r \gg m$. In the limit $r \rightarrow \infty$, the curvature components in these coordinates are given by (with $r_0 = 2m$, and {\it no summation} over repeated indices)
\begin{eqnarray}
R_{0 \mu 0 \mu} &\rightarrow & - \frac{r_0}{r^3}  + O\l( \frac{1}{r^4} \r)  \hspace{2cm} \l({\rm no \,\, summation}\r)
\nonumber \\
R_{0 \mu 0 \nu} &\rightarrow & - \frac{3}{2} \l( \frac{r_0}{r^5} \r) y_{\mu} y_{\nu} + O\l( \frac{1}{r^6}  \r) \;\;\;\;\; (\mu \neq \nu)
\nonumber \\
R_{\mu \nu \mu \nu} &\rightarrow & - \frac{r_0}{2 r^3} + O\l( \frac{1}{r^4} \r) \;\;\;\;\;\;\;\;\;\;\;\;\;\;\;\;\;\;\; (\mu \neq \nu)
\nonumber \\
R_{\mu \nu \mu \lambda} &\rightarrow &  - \frac{3}{2} \l( \frac{r_0}{r^5} \r) y_{\nu} y_{\lambda} + O\l(\frac{1}{r^6}\r) \;\;\;\;\; (\mu \neq \nu \neq \lambda)
\nonumber \\
\label{eq:curvature-isotropic}
\end{eqnarray}

We now wish to compare two situations: an accelerated observer in flat spacetime, and an accelerated observer \textit{with this same acceleration} but in presence of the mass $m$. We expect that such a comparison would provide a natural setting to study what happens when $m$ disappears from the causal domain of the accelerated observer, when it crosses the horizon; in effect, we are accounting for the effect of the particle by considering how the curvature produced by it changes the horizon location (see Figure \ref{fig:rindler-geod}). With this in mind, we can go ahead with the calculation. We are mainly interested in knowing how the horizon location changes when the perturbing mass $m$ moves ``across" the horizon (see Figure \ref{fig:hor-shift}). To lowest order, this can be done by setting $g_{00}=0$, and it is easy to see that one obtains, 
\begin{eqnarray}
z_{0 \pm} = - \frac{1}{a} \l( P \mp \sqrt{P^2 - Q} \r) 
\end{eqnarray}
where
\begin{eqnarray}
P &=& 1 + a^{-1} R_{0 3 0 \chf{A}} y^\chf{A} - a^{-2} R_{0 3 0 3}
\\
Q &=& 1 + R_{0 \chf{A} 0 \chf{B}} y^\chf{A} y^\chf{B} - a^{-2} R_{0 3 0 3}
\end{eqnarray}
From Eqs. (\ref{eq:curvature-isotropic}), we see that in the limit $r \rightarrow \infty$, it is only the $R_{0303}$ term that matters. Also, we make a further {\it assumption} of using the average of the two roots above, $z_{0\, \mathrm{avg}}=-P/a$, to quantify horizon displacement (without this assumption, it is unclear which root to pick, and 
further, it is not clear what terms containing square root of curvature tensor would mean). We then obtain (see Figure \ref{fig:hor-shift})
\begin{eqnarray}
\Delta z_0 &=& z_{\rm f} - z_{\rm i}
\nn \\
&=& -1/a - z_{0\, \mathrm{avg}}
\nn \\
&=& \frac{r_0}{(r a)^3} \Biggl{|}_{r=1/a}
\nn \\
&=& r_0 
\end{eqnarray}
Since $r_0=2m$, we therefore get 
\begin{eqnarray}
m = \frac{\Delta z_0}{2}
\end{eqnarray}

The above result is in exactly similar to the case of particle capture by a Schwarzschild black hole. In fact, for spherically symmetric black holes, the above result is equivalent to attributing an energy $E=r_{\rm h}/2$ to the horizon (with radius $r_{\rm h}$), so that $\Delta E={\Delta r_{\rm h}}/{2}$ is the change in energy when a particle falls into the black hole. Indeed, this is the energy that appears in \cite{TP-thermod}. (In General Relativity, this definition of energy is equivalent to the so called Misner-Sharp energy associated with the black hole.) 
%
%

\noindent {\it Aside:} Note that one can not argue for the above result on dimensional grounds alone, since more than one length scales are involved. In fact, the scaling of $\Delta z_0$ is an outcome of our (admittedly adhoc) choice of $z_{0\, \mathrm{avg}}$. What is remarkable is that, given all the approximations made, we do get the appropriate numerical factor that has been known in the context of black holes in this very simple model.

 \section{Surface term in Fermi coordinates} \label{app:surface-term}
 \addcontentsline{toc}{section}{Appendix}
From the point of view of thermodynamics, it is of interest to evaluate the surface term $P^c$ in the Einstein-Hilbert action
 \begin{eqnarray}
R \sqrt{-g} = \l(\mathrm{bulk~part}\r) - \partial_c P^c
\end{eqnarray}
given by
 \begin{eqnarray}
 P^c &=& \frac{1}{\sqrt{-g}} \partial_b \l[ (-g) g^{bc} \r]
 \nn \\
&=& \sqrt{-g} \left[ g^{c k} \Gamma^{m}_{k m} - g^{i k} \Gamma^{c}_{i k} \right] \label{pc}
 \end{eqnarray}
 Although coordinate dependent, this term can be written in a covariant but observer dependent form, which is the reason why it acquires relevance in the context of the relationship between gravity and thermodynamics. We will calculate this term in the local coordinates based on the wordline of our accelerated observer; such a coordinate system is unique upto general Lorentz transformations on the observer worldline. Since the surface term, to relevant order
 (which we shall make precise soon) is Lorentz invariant, we are effectively probing the space time with the worldlines
 of such observers. Any local information about the space time geometry should be then encoded in the surface term of
 the action (in fact, being made up of second derivatives of the metric, it is only this term that is expected to be relevant in locally inertial coordinates)  \cite{DK-thesis}.
 
 For convenience, we first define three spatial tensors formed from the curvature tensor
 \begin{eqnarray}
 {\mathcal S}_{\alpha \beta} &=& R_{0 \alpha 0 \beta} = {\mathcal S}_{\beta \alpha}
 \\
 {\mathcal E}_{\alpha \beta} &=& (1/4) \varepsilon_{\alpha \gamma \sigma} \varepsilon_{\beta \lambda \mu}
 R_{\gamma \sigma \lambda \mu} = {\mathcal E}_{\beta \alpha}
 \\
 {\mathcal B}_{\alpha \beta} &=& (1/2) \varepsilon_{\alpha \gamma \sigma} R_{0 \beta \gamma \sigma}
 \end{eqnarray}
in terms of which the FNC metric becomes
 \begin{eqnarray}
 {\mathrm d}s^2 = 
 &-& \left[ \left( ~ 1 +  a_{\mu} y^{\mu} ~ \right)^2 + {\mathcal S}_{\mu \nu} y^{\mu} y^{\nu} \right] ~
 {\mathrm d} \tau^2 
 + 2 \left[ - {\tiny{\frac{2}{3}}} \varepsilon_{\rho \alpha \nu} {\mathcal B}_{\rho \mu} y^{\mu} y^{\nu} \right] ~ {\mathrm d} \tau {\mathrm d} y^{\alpha} 
 \nn \\
 &+&
 \left[ \delta_{\alpha \beta} - \frac{1}{3} \varepsilon_{\rho \alpha \mu} \varepsilon_{\sigma \beta \nu} {\mathcal E}_{\rho \sigma} y^{\mu} y^{\nu}  \right] ~
 {\mathrm d} y^{\alpha} {\mathrm d} y^{\beta}
 \end{eqnarray}
 Note that,
 \begin{eqnarray}
 {\mathcal S}^{\alpha}_{~\alpha} = - R^0_{~0} \;, \;\;\; {\mathcal E}^{\alpha}_{~\alpha} = - G^0_{~0}
 \end{eqnarray}
The coordinate system itself is good for length scales $$y \ll {\mathrm{min}} \{a^{-1}, {\mathcal R}^{-1/2}, {\mathcal R} / \partial {\mathcal R}, 1/(a{\mathcal R})^{1/3}\}$$ (symbolically). One can always choose observers for whom the length scale set by acceleration, $a^{-1}$ is much smaller than the curvature dependent terms above. We then
 effectively have a local Rindler observer, with time dependent acceleration, and a horizon determined by $a_{\alpha}(y^{0})$. 

The expression for $P^{\mu}$ in terms of above tensors can be shown to be
 \begin{eqnarray}
P^0 &=& \l(2/3\r) N^{-1} R_{0\mu \nu \mu} y^{\nu} + O_1
\nn \\
P_\mu &=& 2 a_\mu + 2 N^{-1} \mathcal{S}_{\mu \nu} y^\nu + N \l( \mathcal{E}_{\mu \nu} - \mathcal{E}_{\alpha \alpha} \delta_{\mu \nu} \r) y^{\nu} + O_2
\label{eq:Pmu-expr}
\nn \\
 \end{eqnarray}
where $N = 1 + a_\mu y^\mu$, and $O_1, O_2$ represent terms which are higher order in curvature, and/or involve time derivatives of metric components, and/or quadratic in $y^\mu$ \cite{paddy-brazilian-journal}. Note that the terms given above are not necessarily linear in $y^\mu$ (due to the presence of $N$); rather, these are the only terms which can lead to terms linear in $y^\mu$, and hence we have stated them as it is. 
Using identities given above, it is straightforward to see that
 \begin{eqnarray}
\partial_\mu P^\mu  {\Biggl |}_{y^\mu=0} = - R
 \end{eqnarray}
In flat spacetime, $R_{abcd}=0$, and the contribution on any $\tau=\,$constant surface becomes
 \begin{eqnarray}
 \int \DM \tau \int {\mathrm d}^2 y_{_{\perp}} \,\l(2\, n_{\sigma} a^{\sigma}\r)
 \end{eqnarray}
 in obvious notation. For the well known case of a Rindler observer in flat spacetime, this gives the standard contribution of one-quarter of transverse area, when evaluated on the horizon.

\begin{figure}[!htb]
\sidecaption[t]
\scalebox{.4}{\includegraphics{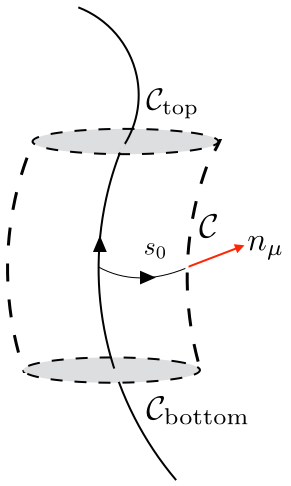}} \hfill
\caption{A tubular neighborhood of the trajectory, with boundary $\mathcal{C} \cup \mathcal{C}_{\rm top} \cup \mathcal{C}_{\rm bottom}$ (see text).}
\label{fig:fnc-modes}
\end{figure}

As an example, let us consider a closed tubular neighbourhood of the trajectory in curved spacetime. That is, at each $\tau$, one sends out geodesics of constant length, say $s_0$, to form a tube, and close this surface at the $\tau=\tau_0$ and $\tau=\tau_0+\Delta \tau$ to obtain a closed surface. The contribution of $P^\mu$ on the curved timelike surface $\mathcal{C}$, which has normal $n^\mu=y^\mu/s_0$, is given (to relevant order) by 
 \begin{eqnarray}
P_\mu n^\mu {\Biggl |}_{\mathcal{C}} = 2 a_\mu n^\mu &+& 2  N^{-1} s_0 \mathcal{S}_{\mu \nu} n^\mu n^\nu 
\nn \\
&+& N s_0 \l( \mathcal{E}_{\mu \nu} - \mathcal{E}_{\alpha \alpha} \delta_{\mu \nu} \r) n^{\mu} n^{\nu}
\nn \\
 \end{eqnarray}
It would be very interesting to explore the detailed mathematical structure of the above expression(s), since it might yield insights into the horizon entropy in curved spacetime.

 \section{Validity of the approximations for the ideal gas calculation} \label{app:idealgas-approxs}
 \addcontentsline{toc}{section}{Appendix}
In this appendix, I give some numerical estimates to illustrate how well the various approximations made in the manuscript hold.

I consider a box of Nitrogen gas, with $m=4.6 \times 10^{-26}$ kg, with approximately $N=6.022 \times 10^{23}$ molecules, at room temperature $k T = 4.11 \times 10^{-21}$ J. I will also take, for the background curvature, the typical magnitude of curvature, say $\mathcal{R}$, produced by the Sun at the location of Earth's orbit \cite{note1}. Since the Sun-Earth distance is $1.496 \times 10^{11}$m and 
Sun's Schwarzschild radius is $3$ k.m., this gives the curvature length scale as 
\begin{eqnarray}
L_{\mathcal{R}} &\approx & \sqrt{(1.496 \times 10^{11})^3 / 3000 {\rm m.}}
\\ \nonumber
&=& 1.056 \times 10^{15} {\rm m.}
\end{eqnarray}
For definiteness, we consider a box of size $L=100$m. In this case, we get the following hierarchy of energy scales
\begin{eqnarray}
\underbrace{E_1 = (m c^2) \times \mathcal{R} \lambda_c^2}_{2.17 \times 10^{-73} {\rm J}} 
\;\; \ll \;\;
\underbrace{E_2 = \frac{\hbar^2}{m L^2}}_{2.42 \times 10^{-47} {\rm J}}
\;\; \ll \;\;
 \underbrace{E_3 = \frac{1}{\beta}=kT}_{4.11 \times 10^{-21} {\rm J}} 
\;\; \ll \;\;
 \underbrace{E_4 = m c^2}_{4.14 \times 10^{-9} {\rm J}}
 \label{eq:energy-estimates}
\end{eqnarray}
%
where $\lambda_c= {\hbar}/{m c} = 7.64 \times 10^{-18}$m is the Compton wavelength. Just a glimpse at these numbers illustrate quite clearly how excellently do the various approximations made in the analysis hold. In fact, one gets a more intuitive understanding of the various numbers and their inter-relationships above by forming their dimensionless ratios.
\begin{eqnarray}
\label{eq:energy-ratios}
{E_1}/{E_2} &=& \mathcal{R} L^2
\\ \nonumber
{E_2}/{E_3} &=& \l({\lambda}/{L}\r)^2
\\ \nonumber
{E_3}/{E_4} &=& \l(\beta m c^2\r)^{-1}
\\ \nonumber
{E_1}/{E_3} &=& \mathcal{R} \lambda^2
\\ \nonumber
{E_2}/{E_4} &=& \l({\lambda_c}/{L}\r)^2
\end{eqnarray}
which, along with the fact that $(\lambda_c/\lambda)^2=(\beta m c^2)^{-1}$, nicely illustrates the self-consistency of the  approximations used, viz:
\begin{enumerate}
\item non-relativistic: $\beta m c^2 \gg 1$
\item validity of Fermi coordinates: $\mathcal{R} L^2 \ll 1$
\item use of Boltzmann distribution: $\lambda \ll L$
\end{enumerate} 
which imply that the first three of the energy ratios (\ref{eq:energy-ratios}) above are {\it small} compared to unity, and the smallness of the last two follow from them. We have therefore shown that our 3 approximations are sufficient to ensure the above mentioned hierarchy of energy scales, which, as illustrated, holds very well for the typical case of $N_2$ gas in a box of size $L=100$m under the assumption that the background curvature is that produced by Sun at location of Earth's orbit. In fact, for our example
\begin{enumerate}
\item $\beta m c^2 = 1.01 \times 10^{12}$
\item $\mathcal{R} L^2 = 8.96 \times 10^{-27}$
\item $\lambda/L = 1.92 \times 10^{-13}$
\end{enumerate} 

\textit{Backreaction due to box contents}:

However, we need to take into account some additional constraints, which turn out to be conceptually trickier and more restrictive. If the box size $L$ is reduced too much, the density of gas inside the box increases, which has following implications: 
\begin{enumerate}
\item The gas can no longer be treated as Maxwell-Boltzmann (as was done in \cite{dk-idealgas}).
\item The curvature produced by the box contents itself might become stronger than the background curvature, in which case the Fermi metric based on background curvature can not be used. And finally, a related fact that, 
\item the energy content of the box might result in a black hole and engulf the box if the Schwarzschild radius $L_{\rm schw}$ corresponding to box's energy content exceeds $L$. 
\end{enumerate}
I discuss the constraints corresponding to (2) and (3) above, and leave (1) for future work since treatment of Bose-Einstein or Fermi-Dirac statistics for this problem requires much further work. Since $m c^2$ is the largest of all energies (per particle), one can use it to make the required estimates. Taking 
\begin{eqnarray*}
L_{\rm schw} \approx \frac{2 G (N m)}{c^2}
\end{eqnarray*}
condition (3) above requires $L_{\rm schw} < L$. On the other hand, Einstein equations imply, for the curvature produced by the box contents, an estimate $\mathcal{R}_{\rm box} \approx L_{\rm schw}/L^3$. To satisfy condition (2), we need $\mathcal{R}_{\rm box} \ll \mathcal{R}$, which becomes equivalent to $L_{\rm schw}/L \ll \mathcal{R} L^2$. Since we require $\mathcal{R} L^2 \ll 1$ for Fermi coordinates based on background curvature to be applicable, the above condition therefore provides a quite stringent upper bound on precisely {\it how small} must $L_{\rm schw}/L$. The physics is quite clear: the smaller the box dimensions, the better the quadratic expansion in Fermi coordinates becomes, but this also increases the density of gas in the box, which might no longer allow the box to be treated as a perturbation over a given, fixed background spacetime. 

However, the above conditions are possible to satisfy, and indeed are satisfied excellently in our example. Following is the hierarchy of length scales which illustrate the numbers involved (all lengths are in meters):
\begin{eqnarray*}
L &=& 10, 100, 1000
\\
L_{\mathcal{R}_{\rm box}} &=& 6.98 \times 10^{15}, 2.21 \times 10^{17}, 6.98 \times 10^{18}
\\
L_{\mathcal{R}} &=& 1.06 \times 10^{15}
\\
\Lambda &=& 7.70 \times 10^{-6}
\\
\lambda &=& 1.92 \times 10^{-11}
\\
\lambda_c &=& 7.64 \times 10^{-18}
\\
L_{\rm schw} &=& 4.11 \times 10^{-29}
\end{eqnarray*}
The magnitudes of $L_{\mathcal{R}_{\rm box}}$ and $L_{\mathcal{R}}$ indicate the sensitivity of the issue of backreaction, discussed above, to the size $L$ of the box. For $L=10$ m., the curvature length scales of the background and the box are of the same order ($L_{\mathcal{R}_{\rm box}} \sim L_{\mathcal{R}}$), hence a larger box is essential for self-consistency of the approximations used. For $L > 100$ m., $L_{\mathcal{R}_{\rm box}} \gg L_{\mathcal{R}}$, and hence one can safely use the Fermi coordinate expansion based on background curvature.


\begin{thebibliography}{99.}
\bibitem{note1} This is simply a choice to provide an estimate of curvature of a typical background spacetime. One could, of course, imagine a more physical situation in which a box is actually present near Earth, in which case the curvature of Earth would be more dominant. In fact, the curvature due to Earth in such a case would be $\sim 10^8$ times that of Sun. However, it is clear from the numerical values in Eq.~(\ref{eq:energy-estimates}) that even this big a factor makes no difference to the arguments!
\end{thebibliography}
\end{document}